\documentclass[12pt]{article}
\usepackage{a4wide}
\usepackage{epsfig}
\usepackage{amsmath}

\newlength{\absize}
\catcode`@=11
\def\citer{\@ifnextchar [{\@tempswatrue\@citexr}{\@tempswafalse\@citexr[]}}

\def\@citexr[#1]#2{\if@filesw\immediate
  \write\@auxout{\string\citation{#2}}\fi
  \def\@citea{}\@cite{\@for\@citeb:=#2\do
    {\@citea\def\@citea{--\penalty\@m}\@ifundefined
       {b@\@citeb}{{\bf ?}\@warning
       {Citation `\@citeb' on page \thepage \space undefined}}%
\hbox{\csname b@\@citeb\endcsname}}}{#1}}
\catcode`@=12

\begin{document}
  \thispagestyle{empty}
  \pagestyle{empty}
  \renewcommand{\thefootnote}{\fnsymbol{footnote}}
\newpage\normalsize
    \pagestyle{plain}
    \setlength{\baselineskip}{4ex}\par
    \setcounter{footnote}{0}
    \renewcommand{\thefootnote}{\arabic{footnote}}
\newcommand{\preprint}[1]{%
  \begin{flushright}
    \setlength{\baselineskip}{3ex} #1
  \end{flushright}}
\renewcommand{\title}[1]{%
  \begin{center}
    \LARGE #1
  \end{center}\par}
\renewcommand{\author}[1]{%
  \vspace{2ex}
  {\Large
   \begin{center}
     \setlength{\baselineskip}{3ex} #1 \par
   \end{center}}}
\renewcommand{\thanks}[1]{\footnote{#1}}
\begin{flushright}
\end{flushright}
\vskip 0.5cm

\begin{center}
{\large \bf Deformed ``Commutative" Chern - Simons System}
\end{center}
\vspace{0.5cm}
\begin{center}
Jian-Zu Zhang
\end{center}
\vspace{0.5cm}
\begin{center}
Institute for Theoretical Physics, East China University of
Science and Technology, \\
Box 316, Shanghai 200237, P. R. China
\end{center}
\vspace{0.5cm}
\begin{abstract}
Noncommutative Chern - Simons' system is non-perturbatively
investigated at a full deformed level. A deformed ``commutative"
phase space is found by a non-canonical change between two sets of
deformed variables of noncommutative space.
It is explored that in the ``commutative" phase space all
calculations are similar to the case in commutative space.
Spectra of the energy and angular momentum of the Chern - Simons'
system are obtained at the full deformed level.
The noncommutative-commutative correspondence is clearly showed.
Formalism for the general dynamical system is briefly presented.
Some subtle points are clarified.
\end{abstract}


\clearpage

{\bf 1  Introduction}

Since the topological Chern - Simons (C-S) field theories proposed
by Deser, Jackiw and Templeton \cite{DJT82}, they have been
extensively investigated in literature. Its interesting
characteristic is that it provides topological mass terms for odd
(space-time) dimensional gauge theories.
There is correspondence between topologically massive
electrodynamics in the Weyl gauge and a model in quantum
mechanics, as explained by Dunne, Jackiw and Trugenberger
\cite{DJT90}, similar dynamical effects of C-S system at quantum
mechanical level were also studied \cite{DJT90,DJ93}.
The Chern-Simons interaction plays a crucial role in the quantum
Hall effect, high $T_c$ superconductivity, cosmic string in planar
gravity, etc.
Their generalization to noncommutative space has been considered.
The noncommutative extension of quantum field theory has attracted
extensively attention in literature \citer{CDS,DN}. In order to
get qualitative understanding of noncommutativity affecting
quantum field theory one tries to understand these effects firstly
in low energy sector at the level of noncommutative quantum
mechanics (NCQM) \citer{CST,JZZ04a}. In literature there has been
a large number of papers dealing with noncommutative
C-S gauge theory \citer{CF,BKSY01}. However, there has been
relatively little work exploring the C-S system at the NCQM level.
NCQM, as the one-particle sector of noncommutative quantum field
theories, can be treated in a more or less self-contained way so
that a more detailed study of noncommutative C-S quantum mechanics
is useful.
Noncommutative C-S system was solved at the NCQM level
\citer{JZZ04a} through the undeformed variables of commutative
space to represent the deformed variables of noncommutative space.
It is interest to clarify whether such a
noncommutative-commutative correspondence can be realized at a
full deformed level through a non-canonical change among two sets
of deformed variables of noncommutative space.

In this paper a noncommutative C-S quantum mechanical model, like
the usual harmonic oscillator, serves as a typical example.
A deformed ``commutative" phase space is found by a non-canonical
change between two sets of deformed variables of noncommutative
space. It essentially defers from the treatment about the
noncommutative-commutative correspondence in literature
where a non-canonical change between a set of deformed variables
of noncommutative space and a set of undeformed variables of
commutative space was considered.
The advantage of the ``commutative" framework is that in which all
calculations in noncommutative space are similar to ones in
commutative space. In the deformed ``commutative" phase space and
deformed ``commutative" Fock space the noncommutative C-S system
is non-perturbatively solved at the full deformed level. Spectra
of its energy and angular momentum are obtained.
The noncommutative-commutative correspondence is clearly showed:
results in commutative space smoothly emerge from ones in
noncommutative space when the limit of vanishing noncommutative
parameters is undertaken, even for cases of noncommutative
versions with expressions where noncommutative parameters appear
in the denominator.
Formalism for the general noncommutative dynamical system is
briefly presented. Finally, some subtle points are clarified.

\vspace{0.4cm}

{\bf 2  The C-S Interactions}

\vspace{0.4cm}

{\bf 2.1  Realization of the C-S Process}

At the quantum mechanical level, a C-S quantum mechanical model
can be constructed as follows. A charged particle of mass $\mu$
and electric charge $q(>0)$ moves in the following external
crossed magnetic and electric fields. The electric field $\bf \hat
E$ acts radially in the $x-y$ plane, $\hat E_i= -\mathcal{E}\hat
x_i,$ $(i=1, 2)$ providing a radial harmonic potential where
$\mathcal{E}$ is a constant. The homogeneous magnetic field $\bf
\hat B$ aligned along the $z-$axis. The vector potential $\hat
A_i$ of ${\bf \hat B}$ is chosen as (Henceforth the summation
convention is used)
$\hat A_i=-B\epsilon_{ij}\hat x_j/2, (i,j=1,2),$
where $\epsilon_{ij}$ is a two-dimensional antisymmetric unit
tensor, $\epsilon_{12}=-\epsilon_{21}=1,$
$\epsilon_{11}=\epsilon_{22}=0$.
The particle's motion is confined to be planar and rotationally
symmetric. The deformed C-S Hamiltonian is represented as
\begin{equation}
\label{Eq:H-2}
\hat H=
\frac{1}{2\mu}(\hat p_i+\frac{1}{2}g\epsilon_{ij}\hat x_j)^2
+\frac{1}{2}\kappa\hat x_i^2= \frac{1}{2\mu}\hat p_i^2
+\frac{1}{2\mu}g\epsilon_{ij}\hat p_i\hat x_j
+\frac{1}{2}\mu\omega^2\hat x_i^2,
\end{equation}
where the constant parameters $g=qB/c$ ($c$ is the speed of light
) and $\kappa=q\mathcal{E}.$ In Eq.~(\ref{Eq:H-2}) the term
$g\epsilon_{ij}\hat p_i\hat x_j/2\mu$ plays a role of realizing
analogs of the C-S theory \citer{DJT90}. The frequency
$\omega=\left[g^2/4\mu^2+\kappa/\mu \right]^{1/2},$ where the
dispersive ``mass'' term $g/2\mu$ comes from the presence of the
C-S term. If NCQM is a realistic physics, low energy quantum
phenomena should be reformulated in this framework. In the above
the noncommutative Hamiltonian (\ref{Eq:H-2}) is obtained by
reformulating the corresponding commutative one,
$H=(p_i+g\epsilon_{ij}x_j/2)^2/2\mu+\kappa x_i^2/2$
where $x_i$ and $p_i$ are the undeformed canonical phase space
variables in commutative space,
in terms of the deformed canonical phase space variables $\hat
x_i$ and $\hat p_i$ in nocommutative space.
%

{\bf 2.2  The Noncommutative Phase Space}

The starting point is the deformed Heisenberg - Weyl algebra. We
consider the case of both position - position noncommutativity
(space-time noncommutativity is not considered) and momentum -
momentum noncommutativity. In this case the consistent deformed
Heisenberg - Weyl algebra is \cite{JZZ04a}:
\begin{equation}
\label{Eq:xp}
[\hat x_{i},\hat x_{j}]=i\xi^2\theta\epsilon_{ij}, \qquad [\hat
p_{i},\hat p_{j}]=i\xi^2\eta\epsilon_{ij}, \qquad
[\hat x_{i},\hat p_{j}]=i\hbar\delta_{ij},\;(i,j=1,2),
\end{equation}
where $\theta$ and $\eta$ are constant parameters, independent of
the position and momentum. Here the noncommutativity of canonical
momenta $\hat p_i$ means the intrinsic noncommutativity. The
scaling factor $\xi=(1+\theta\eta/4\hbar^2)^{-1/2}$ is a
dimensionless constant.

There are different ways to construct creation-annihilation
operators. The deformed annihilation-creation operators $(\hat
a_i$, $\hat a_i^\dagger)$ $(i=1,2)$ at the deformed level which
are related to deformed variables $(\hat x_i, \hat p_i)$ are:
\begin{equation}
\label{Eq:aa+1}
\hat a_i=\sqrt{\frac{\mu\omega}{2\hbar}}\left (\hat x_i
+\frac{i}{\mu\omega}\hat p_i\right),\;
\hat a_i^{\dagger}=\sqrt{\frac{\mu\omega}{2\hbar}}\left (\hat x_i
-\frac{i}{\mu\omega}\hat p_i\right) .
\end{equation}
Equation (\ref{Eq:aa+1}) and the NCQM algebra (\ref{Eq:xp}) show
that the operators $\hat a_{i}^\dagger$ and $\hat a_{j}^\dagger$
for the case $i\ne j$ do not commute. When the state vector space
of identical bosons is constructed by generalizing one-particle
quantum mechanics, because of such a noncommutativity  the
operators $\hat a_1^\dagger\hat a_2^\dagger$ and $\hat
a_2^\dagger\hat a_1^\dagger$ applied successively to the vacuum
state $|0,0\rangle$ [where the definition of the vacuum state
$|0,0\rangle$ is
$\hat a_i|0,0\rangle=0,\;(i=1,2$)]
do not produce
the same physical state,
$\hat a_1^\dagger\hat a_2^\dagger|0,0\rangle\ne\hat
a_2^\dagger\hat a_1^\dagger|0,0\rangle$.
In order to maintain Bose-Einstein
statistics at the non-perturbation level described by $\hat
a_i^\dagger$ the basic assumption is that operators $\hat
a_i^\dagger$ and $\hat a_j^\dagger$ should be commuting. This
requirement leads to a condition between two noncommutative
parameters $\eta$ and $\theta$:
$\label{Eq:dd} \eta=\mu^2\omega^2 \theta.$
From Eqs.~(\ref{Eq:xp}), (\ref{Eq:aa+1})
it
follows that the commutation relations of $\hat a_i$ and $\hat
a_j^\dagger$ read
\begin{equation}
\label{Eq:[a,a+]1} [\hat a_1,\hat a_1^\dagger]=[\hat a_2,\hat
a_2^\dagger]=1, [\hat a_1,\hat a_2]=0;\quad [\hat a_1,\hat
a_2^\dagger] =i\xi^2\mu\omega \theta/\hbar.
\end{equation}
The first three equations in (\ref{Eq:[a,a+]1}) are the same
commutation relations as the one in commutative space.

The last equation in (\ref{Eq:[a,a+]1}) codes effects of spatial
noncommutativity. We emphasize that it is consistent with all
principles of quantum mechanics and Bose - Einstein statistics.

The deformed variables $(\hat x_i, \hat p_i)$ has different
realizations by the undeformed variables $x_i$ and $p_i$
\cite{NP}. We consider the following consistent ansatz of
expansions of $\hat x_{i}$ and $\hat p_{i}$ by $x_i$ and $p_i$:
\begin{equation}
\label{Eq:hat-x-x}
\hat
x_{i}=\xi\left(x_{i}-\frac{\theta}{2\hbar}\epsilon_{ij}p_{j}\right),
\quad \hat
p_{i}=\xi\left(p_{i}+\frac{\eta}{2\hbar}\epsilon_{ij}x_{j}\right).
\end{equation}
where $x_{i}$ and $p_{i}$ satisfy the undeformed Heisenberg - Weyl
algebra in commutative space,
$[x_{i},x_{j}]=[p_{i},p_{j}]=0,\;[x_{i},p_{j}]=i\hbar\delta_{ij}.$

\vspace{0.4cm}

{\bf 2.3  Investigation at the Undeformed Level}

We briefly review the investigation of the deformed C-S
Hamiltonian (\ref{Eq:H-2}) at the undeformed level using
undeformed phase space variables
 $x_i$ and $p_i$ \citer{JZZ04a}. Using (\ref{Eq:hat-x-x}) the C-S
Hamiltonian (\ref{Eq:H-2}) is represented by $x_i$ and $p_i$ as
\begin{eqnarray}
\label{Eq:CS-H1}
\hat H=\frac{1}{2M}\left( p_i+\frac{1}{2}G\epsilon_{ij}
x_j\right)^2 +\frac{1}{2}K x_i^2 =\frac{1}{2M}
p_i^2+\frac{1}{2M}G\epsilon_{ij} p_i x_j+\frac{1}{2}M\Omega^2
x_i^2.
\end{eqnarray}
The above effective parameters $M, G, K$ and $\Omega$ are defined
as
$1/2M\equiv \xi^2\left[c_1^2/2\mu +\kappa
\theta^{2}/8\hbar^2\right],\\
G/2M\equiv\xi^2\left(c_1 c_2/\mu +\kappa \theta/2\hbar\right),
M\Omega^2/2\equiv\xi^2\left[c_2^2/2\mu +\kappa/2 \right],\;
K\equiv M\Omega^2-G^2/4M,$
%
%
where $c_1=1+g\theta/4\hbar,\;c_2=g/2+\eta/2\hbar.$
Equation (\ref{Eq:CS-H1}) is exactly solvable \cite{Baxt,JZZ96}.
We introduce new variables $X_{\alpha}$ and $P_{\alpha}$,
\begin{eqnarray}
\label{Eq:XP}
X_a&=&\sqrt{\frac{M\Omega}{2\omega_a}}x_1-
\sqrt{\frac{1}{2M\Omega\omega_a}}p_2,\quad
X_b=\sqrt{\frac{M\Omega}{2\omega_b}}x_1+
\sqrt{\frac{1}{2M\Omega\omega_b}}p_2,
\nonumber\\
P_a&=&\sqrt{\frac{\omega_a}{2M\Omega}}p_1+
\sqrt{\frac{M\Omega\omega_a}{2}}x_2,\quad
P_b=\sqrt{\frac{\omega_b}{2M\Omega}}p_1-
\sqrt{\frac{M\Omega\omega_b}{2}}x_2,
\end{eqnarray}
where
$\omega_a=\Omega+G/2M,\quad \omega_b=\Omega-G/2M,$
and define the annihilation-creation operators
$A_{\alpha}=\sqrt{\omega_{\alpha}/2\hbar}X_{\alpha}+
i\sqrt{\hbar/2\omega_{\alpha}}P_{\alpha},\;
A_{\alpha}^\dagger=\sqrt{\omega_{\alpha}/2\hbar}X_{\alpha}-
i\sqrt{\hbar/2\omega_{\alpha}}P_{\alpha}, $
$(\alpha=a,b).$ Then the Hamiltonian (\ref{Eq:CS-H1}) decomposes
into two uncoupled harmonic oscillators of unit mass and
frequencies $\omega_a$ and $\omega_b.$
\begin{equation}
\label{Eq:Ha,b} \hat H=H_{a}+H_{b},\;
H_{a,b}=\hbar\omega_{a,b}(A_{a,b}^{\dagger}A_{a,b}+ 1/2),\;
E_{n_a,n_b} =\hbar\omega_a\left(n_a+\frac{1}{2}\right)+
\hbar\omega_b\left(n_b+\frac{1}{2}\right).
\end{equation}

{\bf 3  The ``commutative" phase space}

{\bf 3.1  Construction of Fock space}

The number operators in the hat system are $\hat N_1=\hat
a_1^\dagger\hat a_1$ and $\hat N_2=\hat a_2^\dagger\hat a_2$.
Because the last equation in Eq.~(\ref{Eq:[a,a+]1}) correlates
different degrees of freedom, $\hat N_1$ and $\hat N_2$ do not
commute, $[\hat N_1, \hat N_2]\ne 0.$ They have not common
eigenstates. Thus it is impossible to construct Fock space in the
hat system.
In order to construct a Fock space we introduce the following
auxiliary operators, the tilde annihilation operators $\tilde a_1$
and $\tilde a_2$ \cite{JZZ04a}, and express $\hat a_1$ and $\hat
a_2$ by $\sqrt{2\alpha_1}\tilde a_1$ and $\sqrt{2\alpha_2}\tilde
a_2$ as follows
\begin{equation}
\label{Eq:tilde-a}
%
\hat a_1=\frac{1}{\sqrt{2}} \left(\sqrt{2\alpha_1}\tilde
a_1+\sqrt{2\alpha_2}\tilde a_2\right),\;
\hat a_2=-\frac{i}{\sqrt{2}} \left(\sqrt{2\alpha_1}\tilde
a_1-\sqrt{2\alpha_2}\tilde a_2\right),
\end{equation}
where
\begin{equation}
\label{Eq:alpha-1}
%
\alpha_{1,2}=1\pm \xi^2\frac{\mu\omega\theta}{\hbar}.
\end{equation}
From Eqs.~(\ref{Eq:[a,a+]1}) and (\ref{Eq:tilde-a}) it follows
that the commutation relations of $\tilde a_i$ and $\tilde
a_j^\dagger$ read
\begin{equation}
\label{Eq:tilde[a,a+]}
\left[\tilde a_i,\tilde a_j^\dagger\right]=\delta_{ij},\;
\left[\tilde a_i,\tilde a_j\right]=\left[\tilde a_i^\dagger,\tilde
a_j^\dagger\right]=0,\;(i,j=1,2).
\end{equation}
The algebra (\ref{Eq:tilde[a,a+]}) is same as the bosonic one in
commutative space. The operators $\tilde a_i$ and $\tilde
a_i^\dagger$ are explained as the deformed annihilation and
creation operators in the tilde system. The tilde number operators
$\tilde N_1=\tilde a_1^\dagger\tilde a_1$ and $\tilde N_2=\tilde
a_2^\dagger\tilde a_2$ commute each other, $[\tilde N_1,\tilde
N_2]= 0.$ The commutations between $\tilde a_i$ and $\tilde N_i$
are same as ones in commutative space. The eigenvalues of $\tilde
N_i$ are $n_i=0, 1, 2,\cdots$. A general tilde state is
$\widetilde {|m,n\rangle}\equiv (m!n!)^{-1/2}(\tilde
a_1^\dagger)^m(\tilde a_2^\dagger)^n\widetilde {|0,0\rangle},$
where the vacuum state $\widetilde {|0,0\rangle}$ in the tilde
system is defined as $\tilde a_i\widetilde
{|0,0\rangle}=0,\;(i=1,2).$ It is the common eigenstate of $\tilde
N_1$ and $\tilde N_2$:
$\tilde N_1\widetilde {|m,n\rangle}=m\widetilde {|m,n\rangle}$,
$\tilde N_2\widetilde {|m,n\rangle}=n\widetilde {|m,n\rangle}$,
$(m, n=0, 1, 2,\cdots)$,
and satisfies $\widetilde {\langle m^{\prime},n^{\prime}}
\widetilde {|m,n\rangle}=
\delta_{m^{\prime}m}\delta_{n^{\prime}n}$. The states
$\{\widetilde {|m,n\rangle}\}$ constitute an orthogonal normalized
complete basis of the tilde Fock space. In the tilde Fock space
all calculations are the same as the case in commutative space,
thus the concept of identical particles is maintained and the
formalism of the deformed bosonic symmetry which restricts the
states under permutations of identical particles in multi - boson
systems can be similarly developed.

\vspace{0.4cm}

{\bf 3.2  The ``commutative" phase space}

Starting from the ``commutative" tilde annihilation-creation
operators we introduce the ``commutative" phase space variables
as follows. The tilde annihilation-creation operators $\tilde a_i$
and $\tilde a_i^\dagger$ $(i=1,2)$ and the tilde phase space
variables $\tilde x_i$ and $\tilde p_i$ should satisfy the
following relations:
\begin{equation}
\label{Eq:hat-tilde}
\tilde a_i=\sqrt{\frac{\mu\omega_i}{2\hbar}}\left (\tilde x_i
+\frac{i}{\mu\omega_i}\tilde p_i\right),\;
\tilde a_i^\dagger=\sqrt{\frac{\mu\omega_i}{2\hbar}}\left (\tilde
x_i -\frac{i}{\mu\omega_i}\tilde p_i\right),\;
\omega_i\equiv\alpha_i\omega.
\end{equation}
From Eqs.~(\ref{Eq:tilde-a}) and (\ref{Eq:hat-tilde}) it follows
that $\tilde x_i$ and $\tilde p_i$ satisfy an algebra which is the
same as the undeformed Heisenberg - Weyl algebra in commutative
space:
\begin{equation}
\label{Eq:tilde[xp]}
\left[\tilde x_i,\tilde p_j\right]=i\hbar\delta_{ij},\;
\left[\tilde x_i,\tilde x_j\right]=0,\;\left[\tilde p_i,\tilde
p_j\right]=0,\;(i,j=1,2).
\end{equation}
In view of Eqs.~(\ref{Eq:tilde[xp]}) the tilde phase space can be
considered as ``commutative" one. Calculations in the tilde phase
space are the same as in the commutative phase space. Using
Eqs.~(\ref{Eq:aa+1}), (\ref{Eq:tilde-a}) and (\ref{Eq:tilde[xp]})
we obtain the following relations between two sets of deformed
phase space variables $(\hat x_i$, $\hat p_i)$ and $(\tilde x_i$,
$\tilde p_i)$
\begin{eqnarray}
\label{Eq:tilde-hat}
%
\hat x_1&=&\frac{1}{\sqrt{2}}\left (\alpha_1\tilde x_1+
\alpha_2\tilde x_2\right),\;
\hat x_2=\frac{1}{\sqrt{2}\mu\omega}\left (\tilde p_1-\tilde
p_2\right),
\nonumber\\
\hat p_1&=&\frac{1}{\sqrt{2}}\left (\tilde p_1+\tilde
p_2\right),\quad\quad\;
\hat p_2=-\frac{\mu\omega}{\sqrt{2}}\left(\alpha_1\tilde x_1-
\alpha_2\tilde x_2\right).
\end{eqnarray}
One point that should be emphasized is that {\it the two sets
variables $(\hat x_i$, $\hat p_i)$ and $(\tilde x_i$, $\tilde
p_i)$ are both the deformed phase space variables of
noncommutative space}.
It is worthy noting that the tilde phase space variables $\tilde
x_i$ are always combined with a factor $\alpha_i$, i. e. they are
always represented in the form $\alpha_1 \tilde x_1$ and $\alpha_2
\tilde x_2$, and {\it all} effects of spatial noncommutativity are
included in the parameters $\alpha_i$. Thus results in commutative
phase space smoothly emerge from ones in noncommutative phase
space when the limit of vanishing noncommutative parameters is
undertaken, even for cases of noncommutative versions with
expressions where noncommutative parameters appear in the
denominator (see below).

Like the hat variables $(\hat x_i,\hat p_i)$,  the tilde variables
$(\tilde x_i,\tilde p_i)$ can be expanded by undeformed variables
$(x_i,p_i)$. From Eqs.~(\ref{Eq:hat-x-x}) and (\ref{Eq:tilde-hat})
it follows that
\begin{eqnarray}
\label{Eq:tilde-xp}
\tilde x_1&=&\frac{\xi}{\sqrt{2}\alpha_1}\left [\left (x_1-
\frac{1}{\mu\omega}p_2\right)-\frac{\theta}{2\hbar}p_2+
\frac{\eta}{2\hbar\mu\omega}x_1\right],
\nonumber\\
\tilde x_2&=&\frac{\xi}{\sqrt{2}\alpha_2}\left [\left (x_1+
\frac{1}{\mu\omega}p_2\right)-\frac{\theta}{2\hbar}p_2-
\frac{\eta}{2\hbar\mu\omega}x_1\right],
\nonumber\\
\tilde p_1&=&\frac{\xi}{\sqrt{2}}\left [\left (p_1+\mu\omega
x_2\right)+\frac{\eta}{2\hbar}x_2+
\frac{\theta\mu\omega}{2\hbar}p_1\right],
\nonumber\\
\tilde p_2&=&\frac{\xi}{\sqrt{2}}\left [\left (p_1-\mu\omega
x_2\right)+\frac{\eta}{2\hbar}x_2-
\frac{\theta\mu\omega}{2\hbar}p_1\right].
\end{eqnarray}
Comparing Eqs.~(\ref{Eq:tilde-hat}) and (\ref{Eq:tilde-xp}), it
clearly shows that Eq.~(\ref{Eq:tilde-hat}) represents the
non-canonical changes between two sets of deformed variables
$(\hat x_i,\hat p_i)$ and $(\tilde x_i,\tilde p_i)$, both in
noncommutative space.

\vspace{0.4cm}

{\bf 3.3 Spectra of Energy and Angular Momentum}

Using Eq.~(\ref{Eq:tilde-hat}) the deformed C-S Hamiltonian
(\ref{Eq:H-2}) can be represented by $\tilde x_i$ and $\tilde p_i$
as $\hat H(\hat x_i,\hat p_i)=\tilde H(\tilde x_i,\tilde p_i)$,
here
\begin{eqnarray}
\label{Eq:tilde-H-2}
\tilde H(\tilde x_i,\tilde p_i)&=&\left(\frac{1}{2\mu}\tilde
p_1^2+\frac{1}{2}\mu\omega^2\alpha_1^2\tilde x_1^2\right)+
\left(\frac{1}{2\mu}\tilde
p_2^2+\frac{1}{2}\mu\omega^2\alpha_2^2\tilde x_2^2\right)+
\nonumber\\
&+&\frac{g}{2\mu}\cdot\frac{1}{\omega}\left[\left(\frac{1}{2\mu}\tilde
p_1^2+\frac{1}{2}\mu\omega^2\alpha_1^2\tilde x_1^2\right)-
\left(\frac{1}{2\mu}\tilde
p_2^2+\frac{1}{2}\mu\omega^2\alpha_2^2\tilde x_2^2\right)\right].
\end{eqnarray}
$\tilde H(\tilde x_i,\tilde p_i)$ decouples into two modes of
harmonic oscillators with frequencies $\omega\alpha_1$ and
$\omega\alpha_2$.
In Eq.~(\ref{Eq:tilde-H-2}) all effects of spatial
noncommutativity are included in the parameters $\alpha_1$ and
$\alpha_2$.
Eigenvalues of $\tilde H(\tilde x_i,\tilde p_i)$ can be directly
read out from Eq.~(\ref{Eq:tilde-H-2}):
\begin{equation}
\label{Eq:tilde-E1}
\tilde E_{n_1,n_2} =\hbar\omega\left[\alpha_1
(n_1+\frac{1}{2})+\alpha_2 (n_2+\frac{1}{2}) \right] + \frac{\hbar
g}{2\mu}\left[\alpha_1 (n_1+\frac{1}{2})-\alpha_2
(n_2+\frac{1}{2}) \right]
\end{equation}
Comparing with the results from the investigation at the
undeformed level by using undeformed phase space variables $x_i$
and $p_i$ \cite{JZZ04a}, calculations in the tilde system are
simple. In the limit of vanishing $\theta$ the parameters
$\alpha_i\to 1$ smoothly, Eq.~(\ref{Eq:tilde-E1}) smoothly reduces
to the spectrum in commutative space.
%

The C-S system (\ref{Eq:H-2}) is rotationally symmetric, so the
deformed angular momentum
commutes with the Hamiltonian $\hat H$ in Eq.~(\ref{Eq:H-2}).
In noncommutative space there are different ways to define the
deformed angular momentum. Here we consider the following
definition \cite{JZZ04a}:
\begin{equation}
\label{Eq:Ja}
\hat J_z=\frac{1}{1-\xi^4\theta\eta/\hbar^2}
\left[\epsilon_{ij}\hat x_i\hat
p_j+\frac{\xi^2}{2\hbar}\left(\eta\hat x_i\hat x_i+\theta\hat
p_i\hat p_i\right)\right].
\end{equation}
This deformed angular momentum $\hat J_z$ transforms $\hat x_i$
and $\hat p_j$ as two dimensional vectors:
$[\hat J_z,\hat x_i]=i\epsilon_{ij}\hat x_j,\;[\hat J_z,\hat
p_i]=i\epsilon_{ij}\hat p_j.$
Thus it is a generator of rotations at the deformed level.

Using Eq.~(\ref{Eq:tilde-hat}) the deformed angular momentum $\hat
J_z(\hat x_i,\hat p_i)$ can be reformulated by $\tilde x_i$ and
$\tilde p_i$ as $\hat J_z(\hat x_i,\hat p_i)=\tilde J(\tilde
x_i,\tilde p_i)$, here
\begin{eqnarray}
\label{Eq:J3}
\tilde J(\tilde x_i,\tilde
p_i)&=&\frac{1}{\alpha_1\alpha_2\omega}
\left[\left(\frac{1}{2\mu}\tilde
p_2^2+\frac{1}{2}\mu\omega^2\alpha_2^2\tilde x_2^2\right)-
\left(\frac{1}{2\mu}\tilde
p_1^2+\frac{1}{2}\mu\omega^2\alpha_1^2\tilde x_1^2\right)\right]+
\nonumber\\
&+&\frac{\xi^2\theta\mu}{\alpha_1\alpha_2\hbar}
\left[\left(\frac{1}{2\mu}\tilde
p_1^2+\frac{1}{2}\mu\omega^2\alpha_1^2\tilde x_1^2\right)+
\left(\frac{1}{2\mu}\tilde
p_2^2+\frac{1}{2}\mu\omega^2\alpha_2^2\tilde x_2^2\right)\right]
\end{eqnarray}
The first two terms in Eqs.~(\ref{Eq:tilde-H-2}) and (\ref{Eq:J3})
are the dominate ones. The behavior of $\tilde J(\tilde x_i,\tilde
p_i)$ is different from $\tilde H(\tilde x_i,\tilde p_i)$: the
dominate contribution of $\tilde H$ comes from mode 1 plus mode 2,
but the dominate contribution of $\tilde J$ comes from mode 2
minus mode 1.
Eigenvalues of $\tilde J(\tilde x_i,\tilde p_i)$ can be directly
read out from Eq.~(\ref{Eq:J3}):
\begin{equation}
\label{Eq:J4}
\tilde J_{n_1,n_2} =\frac{\hbar}{\alpha_1\alpha_2} (n_2-n_1) +
\frac{\xi^2\theta\mu\omega}{\alpha_1\alpha_2}(n_1+n_2+1)
\end{equation}
In the limit of vanishing $\theta$ the parameters $\alpha_i\to 1$
smoothly. Though  $\alpha_1$ and $\alpha_2$ appear in the
denominator, Eqs.~(\ref{Eq:J3}) and (\ref{Eq:J4}) smoothly reduces
to the ones in commutative space.
%

The Hamiltonian and angular momentum of the C-S system can be also
reformulated by the tilde annihilation-creation operators $\tilde
a_i$ and $\tilde a_i^\dagger.$ Using Eqs.~(\ref{Eq:xp}) and
(\ref{Eq:aa+1}), the Hamiltonian $\hat H(\hat x,\hat p)$ in
Eq.~(\ref{Eq:H-2}) is first represented by the deformed
annihilation-creation operators $\tilde a_i$ and $\tilde
a_i^\dagger$ as
\begin{equation}
\label{Eq:H1}
\hat H(\hat x,\hat p)=\hbar\omega\left[(\hat a_1^\dagger\hat
a_1+\frac{1}{2})+
(\hat a_2^\dagger\hat a_2+\frac{1}{2})\right]
-i\frac{\hbar g}{2\mu}\left(\hat a_2^\dagger\hat a_1-\hat
a_1^\dagger\hat a_2+
i\frac{1}{\hbar}\xi^2\theta\mu\omega\right).
\end{equation}
Then using Eqs.~(\ref{Eq:tilde-a}) and (\ref{Eq:tilde[a,a+]}), it
can be further represented by $\tilde a_i$ and $\tilde
a_i^\dagger$ as $\hat H(\hat x,\hat p)$= $\tilde H(\tilde a,\tilde
a^\dagger)$, where
\begin{equation}
\label{Eq:tilde-H1}
\tilde H(\tilde a,\tilde a^\dagger)=\tilde
H=\hbar\omega\left[\alpha_1(\tilde a_1^\dagger\tilde
a_1+\frac{1}{2})+\alpha_2(\tilde a_2^\dagger\tilde
a_2+\frac{1}{2})\right]+ \frac{\hbar g}{2\mu}\left[\alpha_1(\tilde
a_1^\dagger\tilde a_1+\frac{1}{2})-\alpha_2(\tilde
a_2^\dagger\tilde a_2+\frac{1}{2})\right]
\end{equation}
Eigenvalues of
$\tilde H(\tilde a,\tilde a^\dagger)$ can be directly read out
from Eq.~(\ref{Eq:tilde-H1}).

Similarly, the deformed angular momentum $\hat J_z$ can be
represented by $\hat a_i$ and $\hat a_i^\dagger$ as
\begin{equation}
\label{Eq:J1}
\hat
J_z=\frac{1}{1-\xi^4\mu^2\omega^2\theta/\hbar^2}\left\{i\hbar\left(\hat
a_2^\dagger\hat a_1-\hat a_1^\dagger\hat
a_2+\frac{i}{\hbar}\xi^2\theta\mu\omega\right)+
\xi^2\theta\mu\omega\left[\left(\hat a_1^\dagger\hat
a_1+\frac{1}{2}\right)+\left(\hat a_2^\dagger\hat
a_2+\frac{1}{2}\right)\right]\right\}.
\end{equation}
Then it can be further represented by $\tilde a_i$ and $\tilde
a_i^\dagger$ as $\hat J_z(\hat a_i, \hat a_i^\dagger)=\tilde
J(\tilde a_i, \tilde a_i^\dagger)$, here
\begin{eqnarray}
\label{Eq:J2}
\tilde J(\tilde a_i, \tilde
a_i^\dagger)&=&\frac{\hbar}{\alpha_1\alpha_2}\left[\alpha_2\left(\tilde
a_2^\dagger\tilde a_2+
\frac{1}{2}\right)-
\alpha_1\left(\tilde a_1^\dagger\tilde a_1+
\frac{1}{2}\right)\right] +
\nonumber\\
&+&\frac{\xi^2\theta\mu\omega}{\alpha_1\alpha_2}\left[\alpha_1\left(\tilde
a_1^\dagger\tilde a_1+\frac{1}{2}\right)+
\alpha_2\left(\tilde a_2^\dagger\tilde
a_2+\frac{1}{2}\right)\right].
\end{eqnarray}
The eigenvalues of $\tilde J(\tilde a_i, \tilde a_i^\dagger)$ can
be directly read out from Eq.~(\ref{Eq:J2}).

\vspace{0.4cm}

{\bf 4  The General Cases}

We briefly express results for a general system. The general
representations of the deformed annihilation-creation operators
$\hat a_i$ and $\hat a_i^\dagger$ fixed by the deformed Heisenberg
- Weyl algebra are \cite{JZZ04a}:
\begin{equation}
\label{Eq:aa+2}
\hat a_i=\sqrt{\frac{1}{2\hbar}\sqrt{\frac{\eta}{\theta}}}\left
(\hat x_i +i\sqrt{\frac{\theta}{\eta}}\hat p_i\right),\;
\hat
a_i^\dagger=\sqrt{\frac{1}{2\hbar}\sqrt{\frac{\eta}{\theta}}}\left
(\hat x_i-i\sqrt{\frac{\theta}{\eta}}\hat p_i\right).
\end{equation}
In order to construct a Fock space we introduce the tilde
annihilation and creation operators $\tilde a_i$ and $\tilde
a_i^\dagger$ through Eq.~(\ref{Eq:tilde-a}),
then introduce the tilde phase space variables $\tilde x_i$ and
$\tilde p_i$ as the follows:
\begin{equation}
\label{Eq:hat-tilde-1}
\tilde
a_i=\sqrt{\frac{\alpha_i}{2\hbar}\sqrt{\frac{\eta}{\theta}}}
\left(\tilde x_i+
\frac{i}{\alpha_i}\sqrt{\frac{\theta}{\eta}}\;\tilde p_i\right),\;
\tilde a_i^\dagger=
\sqrt{\frac{\alpha_i}{2\hbar}\sqrt{\frac{\eta}{\theta}}}
\left(\tilde x_i-
\frac{i}{\alpha_i}\sqrt{\frac{\theta}{\eta}}\;\tilde p_i\right),
%
\end{equation}
where the parameters $\alpha_i$ read
\begin{equation}
\label{Eq:alpha-1}
\alpha_{1,2}=1\pm \xi^2\sqrt{\theta\eta}/\hbar.
\end{equation}
Eq.~(\ref{Eq:tilde-hat}) indicates that $\hat x_i$ and $\hat p_i$
can be expressed as functions of $\alpha_i\tilde x_i$ and $\tilde
p_i$.
From Eqs.~(\ref{Eq:aa+2}), (\ref{Eq:tilde-a}) and
(\ref{Eq:hat-tilde-1}) it follows that $\hat H(\hat x_i,\hat p_i)$
can be represented as a function of $\alpha_i\tilde x_i$ and
$\tilde p_i$:
\begin{equation}
\label{Eq:tilde-H-3}
\hat H(\hat x_i,\hat p_i)=\tilde H(\alpha_i\tilde x_i,\tilde p_i).
\end{equation}

Eq.~(\ref{Eq:tilde-a}) indicates that $\hat a_i$ and $\hat
a_i^\dagger$ can be expressed as functions of $\sqrt{\alpha_i}
\tilde a_i$ and $\sqrt{\alpha_i} \tilde a_i^\dagger$.
Using Eqs.~(\ref{Eq:aa+1}) and (\ref{Eq:tilde-a}), $\hat H(\hat
x_i,\hat p_i)$ can be represented as a function of
$\sqrt{\alpha_i} \tilde a_i$ and $\sqrt{\alpha_i} \tilde
a_i^\dagger$:
\begin{equation}
\label{Eq:tilde-H-3}
\hat H(\hat x_i,\hat p_i)=\tilde H(\sqrt{\alpha_i} \tilde
a_i,\sqrt{\alpha_i} \tilde a_i^\dagger).
\end{equation}
In
$\tilde H(\sqrt{\alpha_i} \tilde a_i,\sqrt{\alpha_i} \tilde
a_i^\dagger)$
we combine $\sqrt{\alpha_i} \tilde a_i$'s and
$\sqrt{\alpha_i} \tilde a_i^\dagger$'s into tilde number operators
$\alpha_i \tilde N_i$'s as possible.
If $\tilde H$ only contains terms of $\alpha_i \tilde N_i$'s, i.
e. $\tilde H$ can be represented as
$\tilde H(\alpha_i \tilde N_i),$
eigenvalues of $\tilde H$ can be directly read out.
In general cases, besides terms of $\alpha_i \tilde N_i$'s,
there are some terms with surplus $\tilde a_i$'s and/or
$\tilde a_i^\dagger$'s,
such as
$\tilde a_i\tilde N_j$'s and/or
$\tilde a_i^\dagger\tilde N_j$'s.
In these cases the states $\widetilde {|m,n\rangle}$ are not
eigenstates of
$\tilde H(\sqrt{\alpha_i} \tilde a_i,\sqrt{\alpha_i} \tilde
a_i^\dagger)$.
Expectations of these terms in the states $\widetilde
{|m,n\rangle}$ are zero,
$\widetilde {\langle m,n}|\tilde a_i\tilde N_j\widetilde
{|m,n\rangle}=
\widetilde {\langle m,n}|\tilde a_i^\dagger\tilde N_j\widetilde
{|m,n\rangle}=0,$
thus only terms of $\alpha_i \tilde N_i$'s contribute to
expectations of
$\tilde H(\sqrt{\alpha_i} \tilde a_i,\sqrt{\alpha_i} \tilde
a_i^\dagger)$.

\vspace{0.4cm}

{\bf 5  Concluding Remarks}

In the tilde phase space $(\tilde x_i,\tilde p_i)$ and tilde Fock
space $(\tilde a_i,\tilde a_i^\dagger)$ all calculations are the
same as the case in commutative space. {\it All effects of spatial
noncommutativity are included in the parameters $\alpha_i$}. The
noncommutative-commutative correspondence is clearly showed:
results in commutative space smoothly emerge from ones in
noncommutative space when the commutative limit $\theta\to 0$ is
undertaken. It is worth noting that even for cases of
noncommutative versions with expressions where noncommutative
parameters appear in the denominator, for example, $\tilde
J(n_1,n_2)$ in Eq.~(\ref{Eq:J4}), the commutative limit keeps
smooth, because in the limit of vanishing $\theta$ the parameters
$\alpha_i\to 1$ smoothly.

\vspace{0.1cm}

It is noticed that in many field theoretical problems, however,
the passage from the noncommutative space to its commutative limit
has not appeared to be smooth \citer{MRS,BDF} because of
noncommutative versions with expressions where noncommutative
parameters appear in the denominator.
Such behavior refers to the UV/IR mixing that arises in loop
calculations in interacting quantum field theories.
Quantum mechanics does not have any ultraviolet divergence
\cite{divergence} and, therefore, one does not expect any
difficulty in taking the limit of vanishing $\theta$.
It is noticed that there is a singularity of the coordinate
transformation in Eq.~(\ref{Eq:hat-x-x}).

The question is whether the limit for the vanishing noncommutative
parameters can be {\it smoothly} undertaken.
Eq.~(\ref{Eq:hat-x-x}) shows that the determinant $\mathcal{R}$ of
the transformation matrix $R$ between $(\hat x_1,\hat x_2,\hat
p_1,\hat p_2)$ and $(x_1,x_2,p_1,p_2)$ is
$\mathcal{R}=\xi^4(1-\theta\eta/4\hbar^2)^2.$
When $\theta\eta=4\hbar^2$, the matrix $R$ is singular. In this
case the inverse of $R$ does not exit.
But this singular condition does not prevent that the vanishing
limit for noncommutative parameters can be smoothly undertaken.
The point is that the noncommutativity of momenta in
Eq.~(\ref{Eq:xp}) means the intrinsic one. All the experiments
show that corrections of spatial noncommutativity, if any, are
extremely small. Therefore, the parameter $\eta$, like the
parameter $\theta$, should be extremely small. But the singular
condition gives
$\eta=4\hbar^2/\theta.$
Thus an extremely small $\theta$ corresponds to an extremely large
$\eta$. It would lead to that corrections from the
noncommutativity of the momenta would be extremely large which
contradicts all the experiments. The above singular condition is
un-physical. For the realistic noncommutative quantum theory
$\theta\eta$ should be the order $o(\theta^2)$, $o(\eta^2)$ and
$o(\theta\eta)$ which must be much less than $ 4\hbar^2$, thus
there is no problem to take the vanishing limit for noncommutative
parameters smoothly.
In fact, the existing upper bounds of $\theta$ and $\eta$ indicate
that the singular condition cannot be met. The existing upper
bounds of $\theta$ and $\eta$ are, respectively,
$\theta/(\hbar c)^2\le (10 \;TeV)^{-2}$ \cite{CHKLO}
and
$|\sqrt{\eta}\,|\le 1 \mu eV/c$ \cite {BRACZ}.
From these upper bounds
it follows that
$\theta\eta=[\theta/(\hbar c)^2](\eta c^2)\hbar^2\ll 4\hbar^2.$
Therefore, in spite of the problems in quantum field theories
because of loop effects, there is no problem in NCQM to have a
smooth limit for the vanishing noncommutative parameters.
This is guaranteed by the parameters $\alpha_i\to 1$ smoothly in
the limit of vanishing $\theta$.

\vspace{0.4cm}

{\bf Acknowledgments}

This work has been supported by the Natural
Science Foundation of China under the grant number 10575037 and by
the Shanghai Education Development Foundation.

\clearpage


\begin{thebibliography}{99}
\bibitem{divergence}
The only divergence that arises in some noncommutative quantum
mechanical models comes from a phase transition where the density
of states diverges and this may happen at a nontrivial value of
$\theta$.
%
It should be point out that, the singular condition
$\theta\eta=4\hbar^2$ of the coordinate transformation in
Eq.~(\ref{Eq:hat-x-x}) is un-physical, therefore, cannot be
related to the phase transition.
\end{thebibliography}
\end{document}